\documentclass[amsmath,amssymb,superscriptaddress,nobalancelastpage,prl,twocolumn]{revtex4-1}

\usepackage{nicefrac}
\usepackage{xfrac}
\usepackage{hyperref}
\hypersetup{colorlinks,linkcolor=blue,urlcolor=blue,citecolor=orange}
\usepackage{siunitx}
\usepackage{xfrac}
\usepackage{natbib}

\usepackage{graphicx}
\usepackage{varioref}
\usepackage{xr-hyper}
\usepackage{xcolor}
\usepackage{colortbl}
\usepackage{hyperref}
\usepackage{ulem}
\usepackage{braket}
\newcommand{\Ca}{Ca$_{1.8}$Sr$_{0.2}$RuO$_4$}

\newcommand{\Sr}{Sr$_{2}$RuO$_4$}
\newcommand{\Cazero}{Ca$_{2}$RuO$_4$}
\newcommand{\tg}{$t_{2g}$}
\newcommand{\dxy}{$d_{xy}$}
\newcommand{\dxz}{$d_{xz}$}
\newcommand{\dyz}{$d_{yz}$}

\begin{document}

 % \title{Strongly correlated electron gas on surface doped \Ca}
 % \title{Heavy fermion in the critical Mott-insulator composition \Ca}
 % \title{Orbital selected heavy-Fermion band in \Ca}
 %\title{Emergence of strong correlations-driven hidden band in \Ca}
 %\title{Non-Fermi liquid Hund’s metal revealed by photoemission spectroscopy}
 %\title{Orbital Selective Breakdown of Conventional Fermi Liquid Quasiparticles}
  \title{Orbitally-Selective Breakdown of Fermi Liquid Quasiparticles in \Ca}
 
\author{D.~Sutter}
\affiliation{Physik-Institut, Universit\"{a}t Z\"{u}rich, Winterthurerstrasse 190, CH-8057 Z\"{u}rich, Switzerland}
     
\author{M.~Kim}
\affiliation{College de France, 75231 Paris Cedex 05, France}
\affiliation{Centre de Physique Th\'{e}orique, Ecole Polytechnique, CNRS, Univ Paris-Saclay, 91128 Palaiseau, France}
  
\author{C.E.~Matt}
\affiliation{Physik-Institut, Universit\"{a}t Z\"{u}rich, Winterthurerstrasse 190, CH-8057 Z\"{u}rich, Switzerland}
  
\author{M. Horio}
\affiliation{Physik-Institut, Universit\"{a}t Z\"{u}rich, Winterthurerstrasse 190, CH-8057 Z\"{u}rich, Switzerland}
 
\author{R.~Fittipaldi}
\affiliation{CNR-SPIN, I-84084 Fisciano, Salerno, Italy}
\affiliation{Dipartimento di Fisica "E.R.~Caianiello", Universit\`{a} di Salerno, I-84084 Fisciano, Salerno, Italy}
  
\author{A.~Vecchione}
\affiliation{CNR-SPIN, I-84084 Fisciano, Salerno, Italy}
\affiliation{Dipartimento di Fisica "E.R.~Caianiello", Universit\`{a} di Salerno, I-84084 Fisciano, Salerno, Italy}
 
\author{V.~Granata}
\affiliation{CNR-SPIN, I-84084 Fisciano, Salerno, Italy}
\affiliation{Dipartimento di Fisica "E.R.~Caianiello", Universit\`{a} di Salerno, I-84084 Fisciano, Salerno, Italy}

\author{K.~Hauser}
\affiliation{Physik-Institut, Universit\"{a}t Z\"{u}rich, Winterthurerstrasse 190, CH-8057 Z\"{u}rich, Switzerland}

\author{Y.~Sassa}
\affiliation{Department of Physics and Astronomy, Uppsala University, S-75121 Uppsala, Sweden}

\author{G.~Gatti}
\affiliation{Institute of Physics, \'{E}cole Polytechnique Fed\'{e}rale de Lausanne (EPFL), CH-1015 Lausanne, Switzerland}

\author{M.~Grioni}
\affiliation{Institute of Physics, \'{E}cole Polytechnique Fed\'{e}rale de Lausanne (EPFL), CH-1015 Lausanne, Switzerland}
    
\author{M.~Hoesch}
\affiliation{Diamond Light Source, Harwell Campus, Didcot, OX11 0DE, United Kingdom}
        
\author{T.~K.~Kim}
\affiliation{Diamond Light Source, Harwell Campus, Didcot, OX11 0DE, United Kingdom}

\author{E.~Rienks}
\affiliation{Helmholtz Zentrum Berlin, Bessy II, 12489 Berlin, Germany}

\author{N.~C.~Plumb}
\affiliation{Swiss Light Source, Paul Scherrer Institut, CH-5232 Villigen PSI, Switzerland}
 
\author{M.~Shi}
\affiliation{Swiss Light Source, Paul Scherrer Institut, CH-5232 Villigen PSI, Switzerland}
    
\author{T.~Neupert}
\affiliation{Physik-Institut, Universit\"{a}t Z\"{u}rich, Winterthurerstrasse 190, CH-8057 Z\"{u}rich, Switzerland}

\author{A.~Georges}
\affiliation{College de France, 75231 Paris Cedex 05, France}
\affiliation{Centre de Physique Th\'{e}orique, Ecole Polytechnique, CNRS, Univ Paris-Saclay, 91128 Palaiseau, France}
\affiliation{Department of Quantum Matter Physics, University of Geneva, 1211 Geneva 4, Switzerland} 
\affiliation{Center for Computational Quantum Physics, Flatiron Institute. 162 5th av. New York NY 10010 USA}
  
\author{J.~Chang}
\affiliation{Physik-Institut, Universit\"{a}t Z\"{u}rich, Winterthurerstrasse 190, CH-8057 Z\"{u}rich, Switzerland}

\begin{abstract}
$~$ \newline
{\textrm{
%Correlated metals are typically classified either as Fermi liquids (FL) or non-Fermi liquids (nFL) depending on whether resistivity scales with temperature squared or not. 
%Transport experiments, however, suggest that
%There is, however, transport evidence suggesting that 
%some materials are hybrids of these two metal classes. This mixed regime is particularly interesting, as it provides insight into how FL break down and the nature of nFL %non-Fermiliquid 
%quasiparticles (QP). 
%Here we show that \Ca\ is such a hybrid metal. 
We present a comprehensive angle-resolved photoemission spectroscopy study of \Ca. Four distinct bands are revealed and along the Ru-O bond direction their orbital characters are identified through light polarisation analysis and comparison to %to density functional- and 
dynamical mean field theory calculations. %Together, they amount to a filling  of about four electrons per ruthenium site.
%Comparison to a tight-binding model, density functional- and dynamical mean field theory calculations allow us to identify their orbital character of these bands. 
Bands assigned to \dxz, \dyz\ orbitals display Fermi liquid behavior with four-fold quasi particle mass renormalization. Extremely heavy Fermions -- associated with a predominantly \dxy\ band character -- are shown to display non-Fermi liquid behavior. We thus demonstrate that \Ca\ is a hybrid metal with an orbitally-selective Fermi liquid quasiparticle breakdown.}}

%An angle resolved photoemission spectroscopy study uncovering
%low-energy electronic structure of \Ca\ is presented.
%Five different Fermi surface sheets are identified and shown to give the expected band filling (4 electrons per Ru site). This observation excludes the orbital selective Mott scenario in which a subset of bands display insulating behavior. A heavy Fermion pocket is found around the zone corner, showing signatures of strong electron correlations beyond expectations of standard Density Functional Theory. Moreover the existence of another $\Gamma$-centered electron pocket is revealed.
%We have studied the low-energy electronic structure of \Ca\ -- the critical composition of the metal-insulator transition in \CSRO\ -- using angle resolved photoemission spectroscopy. We observe three sheets crossing the Fermi level. The orbital characters are identified with Dynamical Mean Field Theory and a simple tight binding model. Furthermore, we exhibit the strong renomalisation $Z_\mathbf{k}$ of a \dxy\ -orbital driven heavy Fermion pocket and the emergence of a $\Gamma$-centered electron pocket due to strong electron correlations. For this composition, our results dismiss the scenario of an orbital-selective Mott phase.
\end{abstract}

\maketitle
  
%\begin{center}
 %   \textbf{I. Introduction}
%\end{center}
%\textit{Introduction:} 
Correlated metals are typically classified either as Fermi liquids or non-Fermi liquids depending on whether resistivity scales with temperature squared or not. There is, however, transport evidence suggesting that some materials are hybrids of these two metal classes~\cite{CooperScience09}. This mixed regime is of particular interest as it provides insight into how Fermi liquids break down and the nature of non-Fermi liquid quasiparticles. In this context,
multi-orbital metallic systems in conjunction with strong Hund's coupling 
and electron correlations are of great conceptual importance~\cite{Georges2013}.
Such Hund's metals are expected to display orbital differentiated quasiparticle (QP) renormalization effects along with magnetic correlations~\citep{MravljePRL2011}.
In the strongly correlated limit, orbitally selective Mott physics (OSMP) has been explored theoretically~\citep{Anisimov2002,KogaPRL2004,Vojta2010,PRB2005Medici,PRB2005Biermann,PRB2005Ferrero}. The concepts of Hund's metals and OSMP have both been applied to describe band structure renormalization effects in pnictide superconductor compounds~\citep{YinNatPhys2011,YiNatCom2015,Science2017Gerber,Science2017Sprau,PRB2010Aichhorn,Science2017Lee}. It remains, however, unclear whether these systems exhibit genuine heavy Fermion and Mott physics. 
In contrast, the oxide compounds LiV$_2$O$_4$ and \Ca\ are multi-orbital systems where the existence of heavy Fermions are clearly demonstrated from specific heat measurements~\cite{NakatsujiPRL03,KondoPRL1997}.
\Ca\ is furthermore in close proximity to a Mott-Hubbard metal-insulator transition~\cite{NakatsujiPRL2004}.
Angle resolved-photoemission experiments (ARPES) on this system have been interpreted in terms of both the Hund's metal and the OSMP scenario~\citep{ShimoyamadaPRL2009,NeupanePRL2009}.
Resistivity and specific heat indicate that the ground state is a Fermi liquid (FL). However, a thermal excitation of just 1\,K turns the system into a non-Fermi liquid (nFL) state~\cite{NakatsujiPRL03}. Here we present a high-resolution ARPES study, demonstrating that \Ca\ is neither a standard Hunds metal nor representing OSMP. In fact, the thermally excited state constitutes an example of a hybrid metal. Along the Ru-O bond direction, bands with \dxz, \dyz\ orbital character display FL behavior whereas \dxy\ dominated bands host nFL QPs. Breakdown of FL QPs are therefore orbitally selective. This physics might apply to other ruthenate systems such as for example Sr$_3$Ru$_2$O$_7$.

\begin{figure*}
    \begin{center}
        \includegraphics[width=1\textwidth]{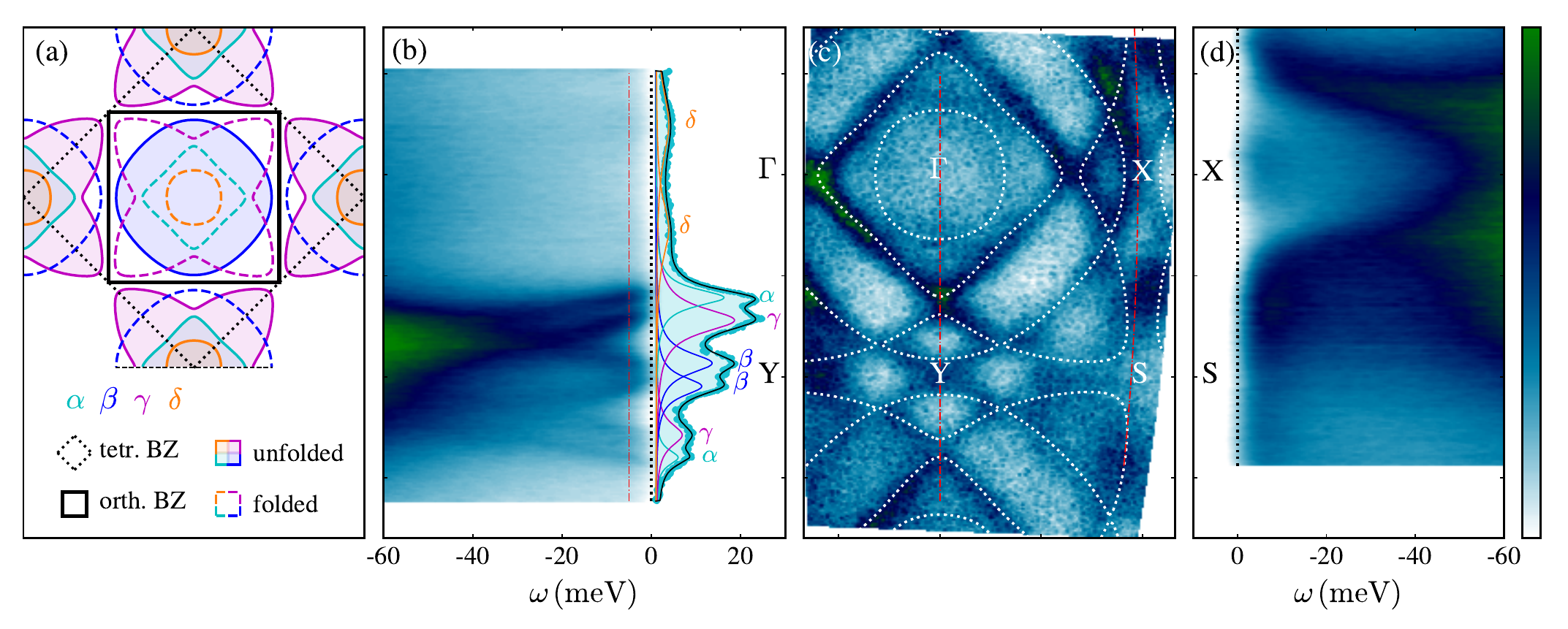}
    \end{center}
    \caption{Low-energy band structure of \Ca. (a) Tight binding (TB) FS (see SM~\cite{SM}, which includes refs.\,(\cite{ZabolotnyyJESRP2013, ChengPRB2010, PuetterPRB2010, NgEPL2000, FatuzzoPRB2015, mizokawaPRL2001, KingmaarXiv2014, WangPRL2004}.)) showing folded (dashed) and unfolded (solid) contours of the $\alpha$, $\beta$, $\gamma$ and $\delta$ sheets. (b) ARPES spectrum along $\Gamma$--Y recorded with 22\,eV circularly polarised light (C$^+$). Cyan circles is an MDC at fixed binding energy $5\pm 1$\,meV, indicated by the red dashed line. The MDC is fitted by eight Lorentzians (total fit in black), displayed colour coded to the corresponding FS sheets in (a). (c) Experimental FS map with the TB model indicated by white dots. The trajectories of ARPES spectra shown (b) and (d) are indicated by dashed red lines. (d) ARPES spectrum along the zone boundary X--S showing flat bands near the Fermi level.}
    %and labelled as indicated.}
    \label{fig:fig1}
\end{figure*}

%\textit{Methods:} %High quality 
Single crystals of \Ca\ were grown by the flux-feeding floating-zone technique~\cite{FukazawaPhysB00,snakatsujiJSSCHEM2001}. ARPES experiments were carried out at  I05, SIS, $1^3$ beamlines of Diamond Light Source (DLS)~\cite{HoeschRevSciInst2017}, Swiss Light Source (SLS), and BESSY -- respectively. %Using a standard top-post~\citep{SutterNatCom2017}, 
All samples were cleaved \emph{in-situ} under UHV conditions and measured at temperatures $T=1-30$\,K. ARPES spectra were collected with different incident photon energies $h\nu$ and light polarisations using Scienta R4000 electron analyzers.
Depending on $h\nu$ and $T$, the overall energy resolution was in the order of 10\,meV. As \Ca\ has  low-temperature L-Pbca~\cite{FriedtPRB2001} crystal structure ($a=5.33$\,\AA, $b=5.32$\,\AA\ and $c=12.41$\,\AA), orthorhombic notation is used. 
The electronic structure is calculated within the DFT+DMFT (density functional theory + dynamical mean field theory) framework using Wien2k~\citep{Blaha2001}  and the TRIQS library~\citep{AichhornPRB2009,AichhornCPC2016, ParcolletCPC2015}, 
including a strong-coupling continuous-time Monte Carlo impurity solver~\citep{GullRMP2011,SethCPC2016}. 
\begin{figure*}
    \begin{center}
 		\includegraphics[width=1\textwidth]{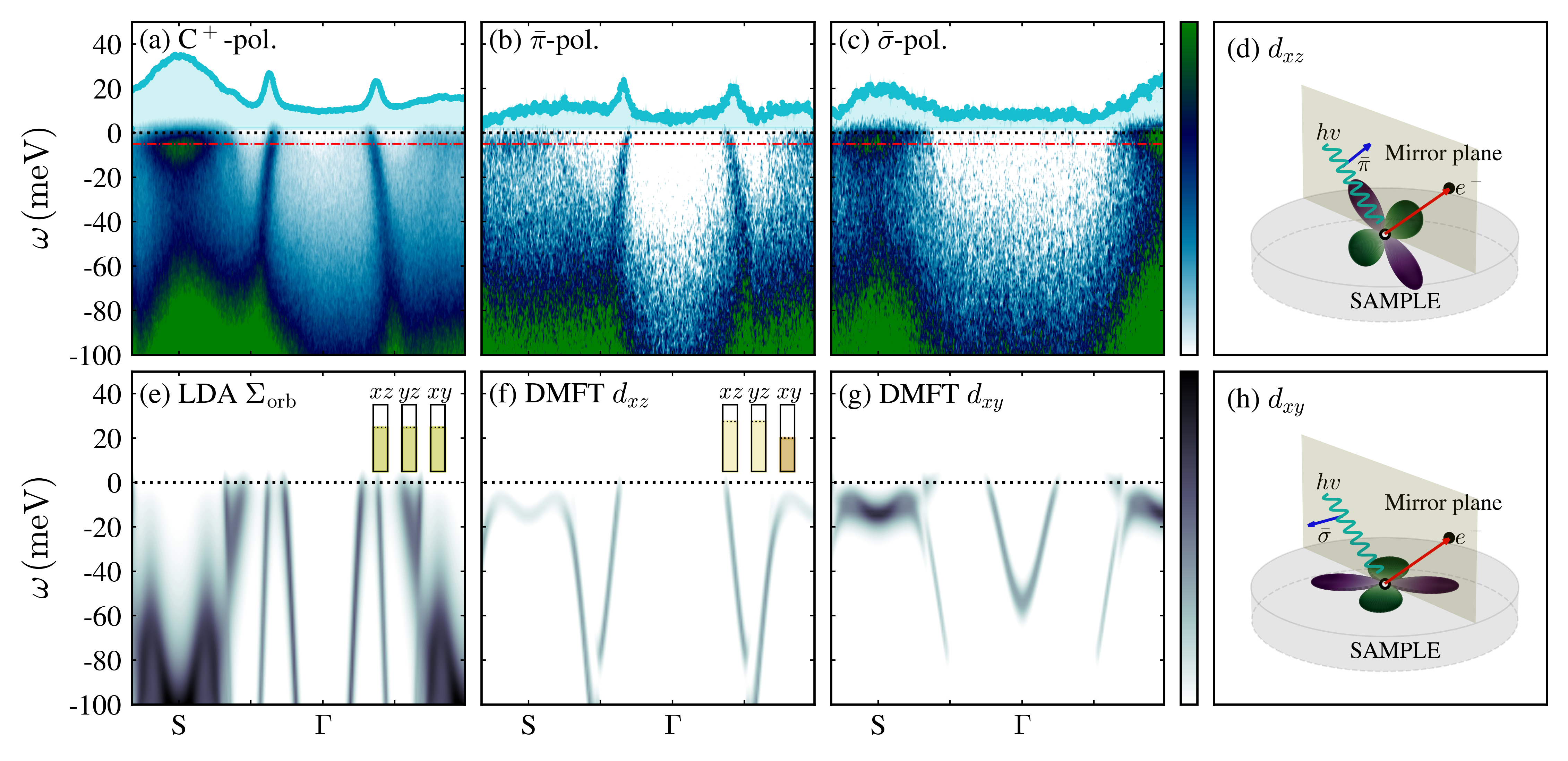}
 	\end{center}
 	\caption{\textrm{Heavy fermion QPs and orbital band character.} (a--c) ARPES spectra along the zone diagonal ($\Gamma$--S) using 40\,eV circularly-, $\bar{\sigma}$-, and $\bar{\pi}$-polarisation, respectively. Cyan points are MDCs near $E_\mathrm{F}$ (dashed turquoise lines). (d,\,h) Schematics for photoemission selection rules for \dxy\ and \dxz\  orbitals. (e) DFT band structure along $\Gamma-$S. (f,\,g) DMFT calculation of the spectral function orbitally resolved. To mimic the experimental data, the DFT and DMFT calculations are plotted in spectral representation, truncated by the Fermi-Dirac distribution ($T_\mathrm{DMFT}=39\,$K), and a constant inverse lifetime of 20\,meV is used. Relative orbital fillings are indicated by the insets in (e,\,f).}
	\label{fig:fig2}
 \end{figure*}
\begin{figure}[h!]
 	\begin{center}
 		\includegraphics[width=0.4\textwidth]{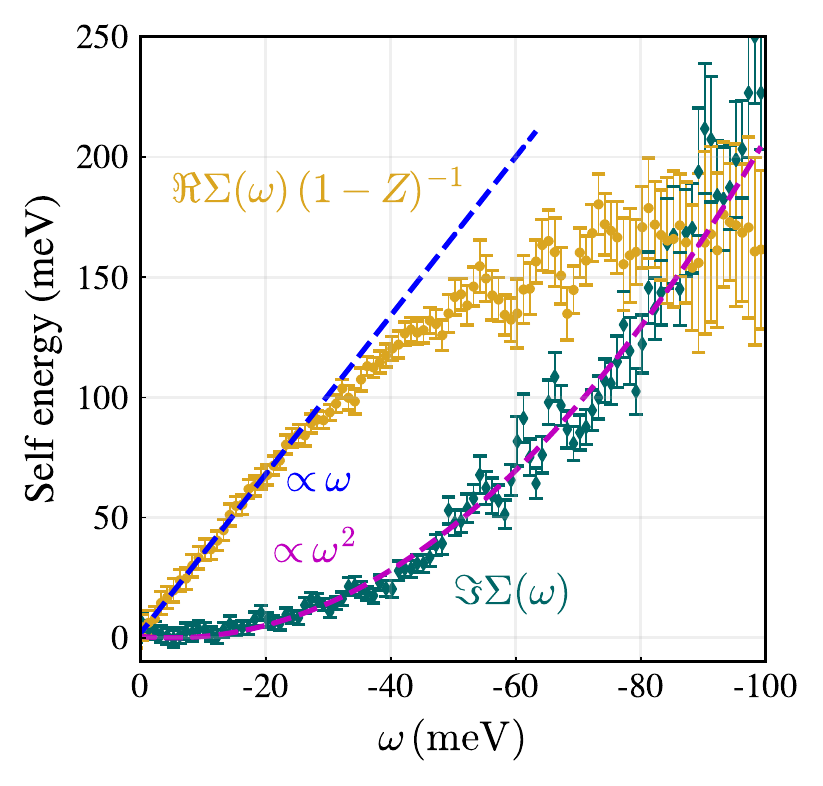}
 	\end{center}
 	\caption{Self-energy $\Sigma(\omega)$ of the $\alpha$-band, plotted as $\Im\Sigma(\omega)$ (green points) and $\Re\Sigma(\omega)/(1-Z)$ (yellow  points) versus binding energy $\omega$. Dashed lines are fits to quadratic and linear dependencies, respectively. For $\Re\Sigma(\omega)/(1-Z)$, the fit is restricted to the low-$\omega$ regime. }
 %	(a) Imaginary (green) $\Im\Sigma$ and real $\Im\Sigma$ Self-energy $\Sigma$ (see text) of the $\bar{\beta}$-band plotted as 
 %	Im$\Sigma$ (black) and Re$\Sigma/ (1-Z)$ (orange) versus $\omega=(E-E_\mathrm{F})$.}  
 	%Corresponding dashed lines 
 	%\textcolor{blue}{are fits using linear $\omega$ ($\chi^2=1.2$) and quadratic $\omega^2$ ($\chi^2=1.1$) dependence.}
 	%(b) Quasiparticle residue $Z$ versus temperature (logarithmic scale). For the $\bar{\beta}$-band, $Z$ is derived from the ratio between observed and DFT Fermi velocities. As discussed in the text, $Z$ is derived from the ratio of coherent and incoherent spectral weight in case of the $\bar{\epsilon}$-band. Finally, the specific heat $C_\mathrm{el}/T$~\citep{BaierPhysB2006,NakatsujiPRL03} is shown converted into $Z$ and plotted versus $T$.}
	 	\label{fig:fig3}
 \end{figure}
 \begin{figure*}[t!]
    \begin{center}
 		\includegraphics[width=1\textwidth]{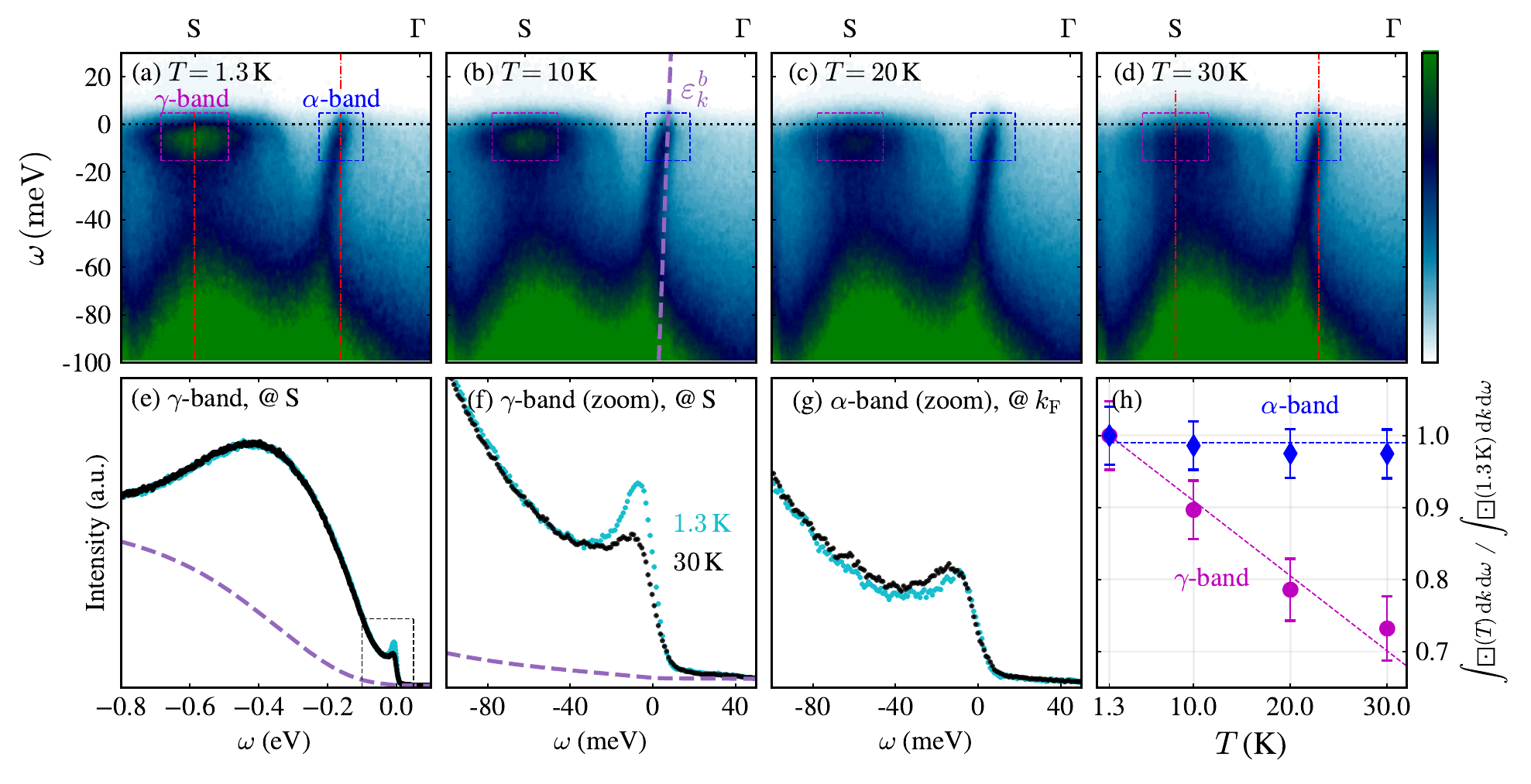}
 	\end{center}
 	\caption{\textrm{Temperature dependence of QP spectral weight.} (a--d) ARPES spectra along S--$\Gamma$ for 
 	 temperatures as indicated with DFT bare band dispersion $\varepsilon_k^b$ in (b). 
 	 %Background subtracted (as in Ref.~\onlinecite{MattNatComm18,HorioNatComm18})
 	 (e--g) Raw data EDCs of the $\alpha$- and $\gamma$-band at $T=1.3$\,K (cyan) and 30\,K (black) and fixed momenta indicated by dashed vertical lines in (a) and (d).  (f) is a zoom near $E_\mathrm{F}$ of the EDC displayed in (e). Dashed lines in (e,f) indicate a Shirley background.
 	 %The shaded area in (e) and (f) represents incoherent and coherent spectral weight modelled by fits to an exponentially modified Gaussian and a Lorentzian function truncated by the Fermi-Dirac distribution, respectively. The dashed line indicates the sum. 
 	 (h) Normalized spectral weight, integrated within the magenta ($\gamma$) and blue ($\alpha$) boxes shown in (a--d), versus $T$.}
	 \label{fig:fig4}
\end{figure*}
%In the DFT part of the computation the Wien2k package~\citep{Blaha2001} was used.
%\textbf{C. Band structure calculations} 
%We calculated the electronic structure within DFT+DMFT using the full potential implementation \citep{AichhornPRB2009} and the TRIQS library~\citep{AichhornCPC2016, ParcolletCPC2015}. In the DFT part of the computation the Wien2k package~\citep{Blaha2001} was used.
%The LDA is used for the exchange-correlation functional. 
%For P
%Projectors on the correlated \tg\ orbital in DFT+DMFT, 
Wannier-like \tg\ orbitals are constructed out of Kohn-Sham bands within the energy window $[-2,1]$\,eV with respect to the Fermi energy $E_\mathrm{F}$. 
For the correct description of atomic multiplets, a %full 
rotationally invariant Kanamori interaction is used~\citep{Georges2013}.
%We used the full rotational invariant Kanamori interaction in order to insure a correct description of atomic multiplets .
%To solve the DMFT quantum impurity problem, we used the strong-coupling continuous-time Monte Carlo impurity solver~\citep{GullRMP2011} as implemented in the TRIQS library~\citep{SethCPC2016}. For the $U$ and $J$ parameters of the Kanamori interaction, we used $U=2.3$\,eV and $J=0.4$\,eV which successfully explains correlated phenomena of other ruthenate such as Sr$_2$RuO$_4$ and $A$RuO$_3$ ($A=$ Ca, Sr) within the DFT+DMFT framework~\citep{MravljePRL2011, DangPRB2015}. 
%We checked that the inclusion of 
Inclusion of charge-self-consistency in the DFT+DMFT loop does not change our results. This validates the correlation induced changes of orbital occupancy in the DFT+DMFT in comparison with the DFT result.

Bulk \Sr\ hosts  three Fermi surface (FS) sheets $\alpha$, $\beta$ (\dxz, \dyz) and $\gamma$ (\dxy)~\citep{DamascelliPRL00,BergemannAiP2003,ZabolotnyyNJP12,HaverkortPRL08,ZabolotnyyJESRP2013,iwasawaPRL2010,IwasawaPRL2012,KondoPRL2016}. Upon Ca for Sr substitution, the $\gamma$-band is undergoing a Liftshitz transition, changing it from electron- to hole-like~\cite{WangPRL2004}. Simultaneously, an electron pocket %(also with \dxy\ character) 
emerging around the zone center is predicted~\cite{KoPRL04}. Orthorhombic folding of these bands (shown schematically in Fig.\,\ref{fig:fig1}a) captures all the observed FS sheets of \Ca. In total four sheets are observed and labelled $\alpha$, $\beta$, $\gamma$ and $\delta$ (Fig.\,\ref{fig:fig1}b--d). 
The weakest $\delta$-band is further documented in the Supplementary Material (SM) SFig.\,1~\cite{SM} and light polarisation dependence of the $\alpha$ and $\gamma$ bands is shown in Fig.\,\ref{fig:fig2}.
The $\alpha$-band, observed with 
C$^+$ and $\bar{\pi}$-polarised light, is suppressed completely in the $\bar{\sigma}$-channel. For the $\gamma$-band in the zone corner, the opposite trend is observed although complete suppression is not found.
Self-energy $\Sigma (\omega)$ versus temperature and binding energy $\omega$ is extracted through a combination of momentum and energy distribution curve (EDC) analysis. %$\Sigma(\omega)$ of the 
For example, the
$\alpha$-band QP dispersion is analyzed by fitting momentum distribution curves (MDC). The resulting band dispersion $\varepsilon_k^\alpha$ and line-width $\Gamma(\omega)$ led us to $\Re\Sigma(\omega)=\varepsilon_k^\alpha-\varepsilon_k^b$ and $\Im\Sigma(\omega)=\Gamma(\omega) v_\mathrm{F}^b$ where $\varepsilon_k^b$ and $v_\mathrm{F}^b$
are the DFT bare band and associated Fermi velocity (Fig.\,\ref{fig:fig3}).  Temperature dependence of spectral intensity along the zone diagonal for both the $\alpha$- and $\gamma$- band are analyzed in Fig.\,\ref{fig:fig4}. In contrast to the $\alpha$-sheet, the $\gamma$-band QP %quasiparticle 
peak amplitude has significant %temperature 
$T$-dependence.

DFT calculations provide an excellent description of the experimental FS of \Sr~\cite{HaverkortPRL08}.
%The agreement is further improved by inclusion of spin-orbit coupling (SOC)~\cite{HaverkortPRL08}.
Already without spin-orbit coupling (SOC), our DFT calculation of \Ca\ produces several of the experimentally observed FS sheets
%as, for example, the $\bar{\alpha}$, $\bar{\beta}$ and $\bar{\delta}$ sheets 
(Fig.\,\ref{fig:fig2}e). SOC is known to improve the calculation along the $\Gamma$--Y direction~\cite{HaverkortPRL08, KimPRL2018}, but has
%should have little or 
no effect along the $\Gamma$--S direction. The absence of the heavy Fermi pocket around the S-point in the DFT calculation %(Fig.\,\ref{fig:fig2}d) 
is therefore a significant discrepancy (compare Fig.\,\ref{fig:fig2}a,\,e).
This motivated our DMFT calculations,  using the same parameters of Coulomb interaction $U=2.3$\,eV and Hund's coupling $J_\mathrm{H}=0.4$\,eV that successfully described \Cazero ~\cite{SutterNatCom2017} and other ruthenates~\cite{DangPRB2015,MravljePRL2011}. 
DMFT predicts strong bandwidth renormalization effects, which is particular clear for the $\delta$-band (see Fig.\,\ref{fig:fig2}g).
Moreover, our DMFT calculation reproduces qualitatively the heavy Fermions states around the zone corner.

Next, we discuss the orbital character of the $\alpha$ and $\gamma$ bands along $\Gamma-$S.
The incident light and centre of our analyser slit, define a mirror plane to which the electromagnetic field has 
odd (even) parity for $\bar{\sigma}$ ($\bar{\pi}$) polarisation (see Fig.\,\ref{fig:fig2}d,h). For final states with even character, selection rules~\cite{DamascelliRMP2003} dictate that
odd (even) band character is suppressed in 
the $\bar{\pi}$ ($\bar{\sigma}$) polarisation channel. The $\alpha$-band being suppressed completely (see Fig.\,\ref{fig:fig2}) in the $\bar{\sigma}$-channel therefore has even character.
Assuming approximately tetragonal crystal structure, (\dxy, \dxz, and \dyz) have (odd, even, and odd) character along the $a$-axis. 
As a result and consistent with the DFT and DMFT calculations, the $\alpha$-band along the Ru-O bond direction has pure \dxz\ character. 
The $\gamma$-band is placed further away from the mirror-plane due to the perpendicular electron analyser-slit.
Hence, less strict selection rules are expected.
Nevertheless, our experimental results and DMFT calculations both assign predominately \dxy\ character to the $\gamma$-band.

As Hund's coupling quenches inter-orbital fluctuations, the orbitals can be viewed approximately as single bands~\cite{LucaPRL2014,MediciPRL2011,Georges2013}.
For the ruthenates, this is valid only along the $\Gamma$-S direction as spin-orbit interaction mixes orbital characters along $\Gamma$-Y~\cite{HaverkortPRL08}. Experimentally, it is thus only sensible to evaluate orbital differentiated properties along $\Gamma$-S.
Comparison of DFT and DMFT suggests that the \dxy\
dominated $\delta$- and $\gamma$-bands are 
most strongly affected by electron correlations.
%The DMFT calculation in addition reveals that these two bands have a significant \dxy\ component (Fig.\,\ref{fig:fig2}e). 
It suggests that electron correlations are orbitally differentiated. 
Orbital fillings provide some insight into this effect. 
As the RuO$_4$ octahedron is almost cubic, DFT yields essentially degenerate \dxy, \dxz, and \dyz\ orbital energies
%. The crystal field splitting between \dxy\ and \dxz, \dyz\ is on the order of 20\,meV 
%and hence the orbital (DFT) 
with equivalent $4/3$-filling (inset Fig.\,\ref{fig:fig2}e). The DMFT calculations, by contrast, indicate that electron interactions favor a less populated \dxy\ orbital with $(n_{xy},n_{xz},n_{yz})=(1.18,1.42,1.42)$ (inset Fig.\,\ref{fig:fig2}f). 
Electron interactions thus push the \dxy\ channel  %system 
closer to half-filling and effectively into a more correlated regime.
%Having an approximately half-filled \dxy-orbital and a van-Hove singularity (vHs) close to the Fermi level, the \dxy-driven bands around the zone center and zone corner ($\delta$ and $\gamma$ sheet) are therefore much stronger renormalized than the \dxz,\dyz\ bands.\\

To describe the orbitally differentiated self-energy of \Ca,
we distinguish between saturated and unsaturated FLs.
The latter refers to QPs for which $Z$ has $\omega$ or temperature dependence. This implies a nFL self-energy, \textit{i.e.}, non-linear $\Re\Sigma(\omega)$ for $\omega \rightarrow 0$.
The saturated regime, by contrast, refers to a standard FL with self-energy $\Sigma(\omega,T)%=\Re \Sigma(\omega) + i\Im \Sigma(\omega, T)
=\gamma_0\omega +i\alpha_0[\omega^2+(\pi k_\mathrm{B}T)^2]$ where $\gamma_0$ and $\alpha_0$ are constants~\cite{VarmaPR2002, DamascelliRMP2003}.
Hence, the QP residue $Z\equiv[1-\partial_\omega \Re \Sigma (\omega)]^{-1}=(1-\gamma_0)^{-1}$ is independent of $\omega$ and
$T$. A FL is therefore expected to display (1) a linear
QP dispersion, (2) a line width that scales as $\omega^2$,  %\jjc{(3) a FL cut-off energy scale $\omega_\mathrm{FL}$ below which $Z|\Im\Sigma(\omega)|<\omega$~\cite{DengPRL2013}
$Z|\Im\Sigma(\omega)|<\omega$~\cite{DengPRL2013} below a cut-off energy scale.
and (4) a QP amplitude proportional to $Z$, independent of $\omega$ and $T$. 
Using $\Re\Sigma(\omega) = \omega\,(1-1/Z)$, the third criterion can be rewritten as $|\Im\Sigma(\omega)|<\Re\Sigma(\omega)/(1-Z)$. For criterion four, the Fermi-Dirac distribution combined with finite instrumental resolution may induce a weak temperature dependence on the effectively observed QP peak amplitude. This weak effect is discussed in the SM Note I~\cite{SM}. 
%For an unsaturated FL, we refer to QPs for which $Z$ has $\omega$ or temperature dependence. This implies a nFL self-energy, \textit{i.e.} non-linear $\Re\Sigma(\omega)$ for $\omega \rightarrow 0$.
Examination of the $\alpha$-band, with pure \dxz, \dyz\ character, reveals an almost $T$-independent QP amplitude (Fig.\,\ref{fig:fig4}g).
%Let's examine the $\alpha$-band that has a pure \dxz\ orbital character (see Fig.\,\ref{fig:fig2}). The observed QP amplitude (proportional to $Z$) is essentially temperature independent (Fig.\,\ref{fig:fig3}g). 
The QP dispersion is approximately linear $\varepsilon_k^{\alpha} \approx v_\mathrm{F}|k-k_\mathrm{F}|$, implying 
%with $v_\mathrm{F}=0.98(6)\,\mathrm{eV}$\AA. 
%With the linearly dispersing bare band ($\varepsilon_\textrm{DFT}^\alpha$ $\approx v_\mathrm{F}^b |k-k_\mathrm{F}|$), this implies that  
$\Re\Sigma(\omega)=(1-v_\mathrm{F}^b/v_\mathrm{F})\,\omega$, with $v_\mathrm{F}$ and $v_\mathrm{F}^b$ being dressed and bare Fermi velocities~\cite{FatuzzoPRB2014}.
Assuming an isotropic FL and using $v_\mathrm{F}^b=2.34\,$eV\AA, the  QP residue yields
$Z= v_\mathrm{F}/v_\mathrm{F}^b = 0.26(4)$, 
consistent with DMFT that finds $Z_{xz}=0.23$.
%Within experimental error bars, $Z$ is therefore temperature independent (Fig.\,\ref{fig:fig4}b). 
%The extracted value of $Z$ is in quantitative agreement with DMFT calculations that finds
%$Z_{xz}=0.23$ ($Z_{yz}=0.21$).
Analysis of the MDC linewidth (HWHM) at $T=30$\,K yields $\Gamma(\omega)= \Gamma_0+\eta \omega^2$ with $\Gamma_0= 0.020(2)\,$\AA$^{-1}$ and $\eta= 10.6(6)\,$\AA$^{-1}$eV$^{-2}$ being constants. This is documented by plotting $\Im\Sigma(\omega)=(\Gamma(\omega)-\Gamma_0)\,v_\mathrm{F}^b$ versus $\omega$ (Fig.\,\ref{fig:fig3}). By comparing $\Re\Sigma(\omega) / (1-Z)$ and $\Im\Sigma(\omega)$, 
%we estimate a  $\omega_\mathrm{FL}\sim 80$\,meV Fermi liquid cut-off  -- in agreement with our DMFT results at $T=39$\,K (see Supplementary Fig.\,2).
criterion three is obviously satisfied and as shown in Fig.~\ref{fig:fig4}g,h the quasiparticle amplitude is temperature independent.
%(Fig.\,\ref{fig:fig4}a). 
%This cut-off energy scale is also found in our DMFT calculations ($T=39$\,K) of the \dxz, \dyz\ orbitals assigned to the $\alpha$-band -- see Supplementary Figure\,1.
The QP excitations of the $\alpha$-band thus fulfil, in the most strict sense, all criteria of a FL (see also SM SFig.\,5~\cite{SM}).
 
Resistivity and specific heat measurements, however, display FL behavior for $T<1$\,K only and much heavier QP masses~\cite{NakatsujiPRL03}. 
Reconciliation is reached by analysis of the extremely dressed $\gamma$-band QP states around the S-point. These QP amplitudes are roughly proportional to $Z$. %, are accessed through EDC's. 
In contrast to the $\alpha$-band, the QP peak amplitude of the $\gamma$-band exhibits a pronounced suppression with increased $T$ (Fig.\,\ref{fig:fig4}e--g). To circumvent the effects of (i) the Fermi-Dirac distribution, (ii) impurity scattering  and (iii) finite instrumental resolution (see SM SFig.\,3 and 4~\cite{SM}), it is useful to perform a box integration of spectral weight around $k_\mathrm{F}$ (see Fig.\,4a--d). Again, the $\gamma$-band displays a pronounce spectral weight temperature dependence whereas the $\alpha$-band remains approximately unchanged. %A box integration of spectral weight around the respective $k_\mathrm{F}$'s reveals the same trends (Fig.\,\ref{fig:fig4}h). 
As both the $\alpha$- and $\gamma$-bands are measured simultaneously, this effect is not a result of surface degradation. We are thus led to conclude that the \dxy\ dominated $\gamma$-band states display non-saturated FL behavior. Furthermore, the ratio between coherent and incoherent spectral weight (see Fig.\,\ref{fig:fig4}e) indicates that $Z\ll1$ around the S-point, in accordance with the DMFT value  $Z_{xy} \approx 0.05$. We have thus demonstrated that the QP mass renormalization and FL QP breakdown are orbitally selective along the $\Gamma$-S direction. It is also worth noticing that temperature dependent spectral weight has also been reported in CeCoIn$_5$~\cite{ChenPRB2017} and Ce$_2$PdIn$_8$~\cite{YaoPRB2019}. This effect may therefore be generic to heavy fermion quasiparticles.

In summary, we have presented a combined ARPES, DFT, and DMFT study of \Ca. Our results revealed the complete low-energy electronic structure. Through light polarisation analysis and band structure calculations, insight into the orbital band character was obtained. By studying self-energy effects, it was demonstrated that QP masses and the FL breakdown are orbitally selective. \Ca\ thus constitutes a unique example of a hybrid metal hosting orbitally differentiated FL and nFL QPs. As an outlook, it is interesting to consider the idea that nFL behavior found in Ba$_2$RuO$_4$~\cite{BurganovPRL2016} and Sr$_2$RuO$_4$~\cite{BarberPRL2018} under strain 
has a similar underlying origin.\\

D.S., M.H., T.N, and J.C. acknowledge support by the Swiss National Science Foundation and Y.S. was supported by the Wenner-Gren foundation. 
Experiments were carried out on the I05, SIS, and $1^3$ endstations at the Diamond Light Source, Swiss Light Source and BESSY, respectively. We acknowledge Diamond Light Source for time on beamline I05 under proposal SI15296.
A.G. and M.K. acknowledge the support of the European Research Council grant ERC-319286-QMAC and the Swiss National Science Foundation (NCCR MARVEL), as well as support from the CPHT computer team. The Flatiron Institute is supported by the Simons Foundation. 
We thank all beamline staff for technical support.\\

\bibliography{CSRO20}
\clearpage

\end{document}